\documentclass[letterpaper, 10pt, conference]{ieeeconf}  

\IEEEoverridecommandlockouts                              
\usepackage{graphics} 
\usepackage{epsfig} 
\usepackage{amsmath} 
\usepackage{amssymb}  
\usepackage{cite}
\usepackage{comment}
\usepackage{xcolor}

\title{\LARGE \bf Observability-Blocking Controls for Double-Integrator and Higher Order Integrator Networks}

\author{Joseph D. Tran$^{1}$ and Abdullah Al Maruf$^{2}$
     \thanks{$^1$California State University, Los Angeles (Cal State LA).
        {\tt\small jtran233@calstatela.edu}}
     \thanks{$^2$California State University, Los Angeles (Cal State LA).
        {\tt\small amaruf@calstatela.edu}}	
	}

\begin{document}

\maketitle

\begin{abstract}

The design of state-feedback controls to block observability at remote nodes is studied for double integrator network (DIN) and higher order integrator network models.
A preliminary design algorithm is presented first for DIN that requires $m+2$ actuation nodes to block observability for the measurement obtained from a set of $m$ nodes. The algorithm is based on eigenstructure assignment technique and leverages the properties of the eigenvectors in DIN. Next, the topological structure of the network is exploited to reduce the number of controllers required for blocking observability. The number of actuation nodes in sparser design depends on the cardinality of a cutset separating the actuation and measurement locations. Later, the design principles are generalized for blocking observability in $N$-th order integrator network models.


\end{abstract}


\section{Introduction}

Controllability and observability of
dynamical networks have been extensively studied in the controls-engineering community \cite{rahmani2009controllability,liu2011controllability,sundaram2012structural,pasqualetti2014controllability,summers2014optimal,wang2016controllability,dhal2015vulnerability,summers2015submodularity,roy2016sensor,li2020structural}. One key question that has found attention recently is how to design controls in a network to shape observability and controllability while maintaining network performance. Such design problems are particularly relevant in contexts where multiple stakeholders (including adversaries) have access to the network's dynamics, and have conflicting objectives in modulating the dynamics. For example, increasing concern about cyber-attacks in the power grid has led to significant interest in designing wide-area control systems that prevent adversaries from estimating or manipulating the dynamics \cite{sridhar2011cyber}. Similarly, controller designs for multi-vehicle systems require security from intruders probing a subset of vehicles to infer the system’s state \cite{xue2014security}. In this regard, observability-blocking has found usefulness in protecting privacy in dynamical networks \cite{zhang2023observability}. 

Based on this motivation, in our previous work we explored design of feedback controls in a linear synchronization network to block observability/controllablity at a remote set of nodes, while maintaining all the eigenvalues and most of the eigenvectors \cite{al2019observability,al2021using,al2022observability}. However, the synchronization model considered in our earlier works has a limitation that each node is associated with only a scalar state. In contrast, double-integrator networks (DIN) and higher-order integrator network dynamics, which are widely used to model and control high-fidelity multi-agent systems such as robotic motion control and tracking in unmanned vehicle networks, consider network nodes with multiple states \cite{cao2010distributed,wang2024time}. Specifically, DIN is extensively used as a network model for the formation control of unmanned vehicle networks, among many other applications \cite{cao2010distributed,xue2014security}. Therefore, in this work we focus on the design of observability blocking controls for DIN model. We also study generalization of our design for higher order integrator networks.

Our main focus of this study is to utilize the dynamics and topological structure of an integrator network to develop a design algorithm that can prevent observability of the measurements obtained from a given set of nodes. Our main contributions for this paper are the following:

1) An algorithm that blocks observability of a set of $m$ nodes in DIN is presented using state-feedback controls at another set of $q=m+2$ nodes. The algorithm builds upon the design from our earlier work \cite{al2022observability} and maintains all the eigenvalues and most of the eigenvectors. Unlike previous work, the algorithm leverages a special property of eigenvectors in a DIN model to limit the actuation to $q=m+2$.

2) A sparser design scheme is then presented that exploits the topological structure of the DIN model. Similar to our previous work, we show that the number of actuation can be reduced by blocking observability at cutsets separating the actuation and measurement locations. However, the conditions required for this design in the context of DIN differ significantly from those established in our earlier work.

3) We then extend our results for $N$-th order integrator networks. The number of actuation nodes required for blocking observability remains the same as in the DIN case.

The paper is organized as follows: In section II, we formulate the observability-blocking controller design problem for DIN. We present our main results on observability-blocking controls in Section III for both DIN and $N$-th order integrator networks. Section IV is devoted to a numerical example, and in Section V we draw conclusions.

\section{Problem Formulation}


In this study, we first focus on a double-integrator network (DIN) model which will allow us develop design principles for higher order integrator network models. Specifically, here we consider a DIN with $n$ nodes labeled as $1, 2, \cdots, n$. Each node $i$ is associated with two scalar states denoted as $s_i$ and $\dot{s}_i$. In the context of a vehicle network, $s_i$ and $\dot{s}_i$ correspond to the position and velocity of the vehicle $i$. The interactions between different nodes in the DIN are represented by a directed graph $\mathcal{G}(\mathcal{V},\mathcal{E})$, where $\mathcal{V}$ contains $n$ nodes/vertices and $\mathcal{E} \subset \mathcal{V} \times \mathcal{V}$ contains a collection of directed edges specified as an ordered pair of nodes. We assume the graph is strongly connected and use $\mathcal{N}(i)$ to denote the set of nodes such that $(j,i)\in \mathcal{E}$. In our DIN model, the node dynamics of each node $i$ is given by \cite{zhang2019double}:
\begin{eqnarray} \label{eq:scalar_dynamics}
\ddot{s_i} &=& -\sum_{j \in \mathcal{N}(i)}{w_{ji}^s(s_{i}-s_{j}}) - \sum_{j \in \mathcal{N}(i)}{w_{ji}^{\dot{s}}(\dot{s}_{i}-\dot{s}_{j}})+u_i \nonumber \\
\end{eqnarray}
Here, weights $w_{ji}^s \in \mathcal{W}^s$ and $w_{ji}^{\dot{s}}\in \mathcal{W}^{\dot{s}}$ quantify the influence of node $j$ on node $i$ through the edge $(j,i)\in\mathcal{E}$ and $u_i$ denotes the actuation applied to the node $i$. In our model, $w_{ji}^s$ and $w_{ji}^{\dot{s}}$ are positive and assumed to not be related for generality. As standard, in order to denote the dynamics conveniently, we consider two (asymmetric) Laplacian matrices $\mathbf{L}^s$ and $\mathbf{L}^{\dot{s}}$ on the graph $\mathcal{G}(\mathcal{V},\mathcal{E})$ for the edge-weight sets $\mathcal{W}^s$ and $\mathcal{W}^{\dot{s}}$. Specifically, $\mathbf{L}^s$ is an $n \times n$ matrix whose entries are as follows: each off-diagonal entry $L_{ij}^s$ is equal to $-w_{ij}$ if $(i,j) \in \mathcal{E}$ and otherwise is set to $0$; each diagonal entry $L^s_{ii}$ is equal to $-\sum_{j=1, j \neq i}^{n} L^s_{ij}$. We define $\mathbf{L}^{\dot{s}}$ similarly based on the edge weight set $\mathcal{W}^{\dot{s}}$. We further define the network state $\mathbf{x}(t)$ as $\mathbf{x}(t)= [s_1(t) ~~ s_2(t) ~ \cdots ~ s_n(t) ~~ \dot{s}_1(t) ~~ \dot{s}_2(t) ~ \cdots ~ \dot{s}_n(t)]^T$. We enhance this network model by capturing a set of nodes as {\em actuation nodes} where actuation can be provided by the network operator (i.e. $u_i \neq 0$ in eqn. (\ref{eq:scalar_dynamics})) and another distinct set of nodes as {\em measurement nodes} that can be accessed by an adversary. We assume there are $q$ actuation nodes and $m$ measurement nodes given by the set $\{r_1, r_2, \hdots, r_q\}$ and set $\{ r'_1, r'_2, \hdots, r'_m \}$, respectively. It is reasonable to assume that when an adversary finds access to a node $r'_i$, where $i=1, 2, \cdots, m $, it can measure both states $s_{r'_i}$ and $\dot{s}_{r'_i}$. 


The dynamics of the DIN with actuation and measurement included are then given by the state space model:
\begin{subequations} \label{dynamics_sub}
\begin{align}
\mathbf{\dot{x}} =&  \begin{bmatrix}
    \mathbf{0} ~~~~~~\mathbf{I} \\ 
    -\mathbf{L}^s ~~ -\mathbf{L}^{\dot{s}} \end{bmatrix} \mathbf{x} + \begin{bmatrix}
    \mathbf{0} \\
    \mathbf{\hat{B}}\end{bmatrix}  \mathbf{u}, \\
\mathbf{y} =& \begin{bmatrix}
    \mathbf{\hat{C}} ~~ \mathbf{0}\\
    \mathbf{0} ~~ \mathbf{\hat{C}}
\end{bmatrix} \mathbf{x}
\end{align}
\end{subequations}
where $\mathbf{\hat{B}}=[\mathbf{e}_{r_{1}} ~ \mathbf{e}_{r_{2}}\cdots ~\mathbf{e}_{r_{q}}]$, $\mathbf{\hat{C}}=[\mathbf{e}_{r'_{1}} ~ \mathbf{e}_{r'_{2}}\cdots ~\mathbf{e}_{r'_{m}}]^T$ and $\mathbf{e}_i \in \mathbb{R}^n$ is a 0--1 indicator vector with $i$ entry equal to $1$. Note, in the above, $\mathbf{u} \in \mathbb{R}^q$ and $\mathbf{y}\in \mathbb{R}^{2m}$ denotes the inputs applied at the actuation nodes and measurements taken at the measurement nodes, respectively. Writing $\mathbf{A} = \begin{bmatrix}
    \mathbf{0} ~~~~~~ \mathbf{I} \\ 
    -\mathbf{L}^s ~~ -\mathbf{L}^{\dot{s}} \end{bmatrix}$, $\mathbf{B} = \begin{bmatrix}
    \mathbf{0} \\
    \mathbf{\hat{B}}\end{bmatrix}$ and $\mathbf{C}=\begin{bmatrix}
    \mathbf{\hat{C}} ~~ \mathbf{0}\\
    \mathbf{0} ~~ \mathbf{\hat{C}}
\end{bmatrix}$, we can write the dynamics in (\ref{dynamics_sub}) as
\begin{subequations} \label{eq: open-loop}
\begin{align} 
\mathbf{\dot{x}} =& \mathbf{A} \mathbf{x} + \mathbf{B} \mathbf{u},\\
\mathbf{y} =& \mathbf{C} \mathbf{x}
\end{align}
\end{subequations}
We assume throughout our development that the pair $(\mathbf{A},\mathbf{B})$ is controllable.
In this study, we consider the design of linear feedback controllers at the actuation nodes, to block the observability of the network dynamics with respect to the measurements $\mathbf{y}$ obtained at the measurement nodes.  As a nominal case, a static state feedback control scheme is considered, with the input at each node $r_i$, $i=1, 2, \hdots, q$, specified as
$u_{r_i}=\mathbf{k}_{r_i}^T\mathbf{x}$ where $\mathbf{k}_{r_i}$ is the control gain. Assembling
the state feedback models for each actuation node and writing $\mathbf{F}=[\mathbf{k}_{r_1} ~\mathbf{k}_{r_2} \cdots \mathbf{k}_{r_q}]^T$, we get
\begin{equation}
\mathbf{u}=\mathbf{F}\mathbf{x}
\end{equation}
Therefore, the closed-loop dynamics for the feedback control become:
\begin{subequations}\label{eq:main}
\begin{align}
\mathbf{\dot{x}}=& (\mathbf{A}+\mathbf{B}\mathbf{F})~\mathbf{x}, \\
\mathbf{y} =& \mathbf{C}\mathbf{x}. 
\end{align}
\end{subequations}

Our focus here is to design the state-feedback controller, defined by the gain matrix $\mathbf{F}$, to enforce that the pair $(\mathbf{C},(\mathbf{A}+\mathbf{BF}))$ is unobservable, while maintaining as much of the open-loop eigenstructure as possible.  
Our main objective is to find sparse feedback controllers that only require a small number actuation by exploiting the dynamics and graph topology of the DIN. We also aim to generalize our result for $N$-th order integrator networks.



\section {Main Results}


In this section, we present our results for the design of observability-blocking controls. First, we develop a design algorithm for DIN requiring $q=m+2$ actuation nodes. Then, we provide a sparser design that utilizes the topological structure of DIN to reduce the required number of actuation nodes. Finally, we extend our results to encompass $N$-th order integrator networks.

\subsection{General Design of Observability-Blocking Controller for DIN}

In this section, we present a general algorithm for designing a state-feedback controller that makes the pair $(\mathbf{C},(\mathbf{A}+\mathbf{BF}))$ unobservable while preserving much of the open-loop eigenstructure. The design relies on the eigenstructure assignment method, where a selected eigenvector will be modified to enforce unobservability. We stress that the algorithm presented here is built upon our previous algorithm (see Algorithm 1 in \cite{al2022observability}). Our main contribution for this section is that we provide a generalization of the design algorithm given in \cite{al2022observability} so that it can be applied to a DIN while keeping the required actuation the same as before.  

For the simplicity of our development, we assume eigenvalues are non-defective (i.e. algebraic and geometric multiplicities of each eigenvalue are same). However, the result also applies for non-defective eigenvalues as discussed later. We label the eigenvalues of $\mathbf{A}$ as $\lambda_1, \lambda_2, \hdots, \lambda_{2n}$, and note that corresponding eigenvectors $\mathbf{v}_1, \mathbf{v}_2, \hdots, \mathbf{v}_{2n}$ can be found which are linearly independent and span $\mathbb{C}^{2n}$. In the ensuing development, for notational convenience and without loss of generality, the last $m$ nodes (nodes $n-m+1, \hdots, n$) are assumed to be the measurement nodes. Hence, $\hat{\mathbf{C}}=[\mathbf{e}_{n-m+1}~\mathbf{e}_{n-m+2} ~\cdots ~\mathbf{e}_n]^T$. 

The design algorithm is based on selecting one of the eigenvalues of $\mathbf{A}$ and its associated eigenvector, say $\lambda_p$ and $\mathbf{v}_p$ where $p \in \{1, 2, \hdots, 2n \}$. A feedback controller is constructed using eigenstructure assignment method, such that the eigenvector $\mathbf{v}_p$ is modified to a vector $\hat{\mathbf{v}}_p$ whose entries corresponding to the measurement nodes i.e. $n-m+1, n-m+2, \cdots, n$ and $2n-m+1, 2n-m+2, \cdots, 2n$ entries are all zeros ($\bar{\mathbf{v}}_p$ is also modified commensurately if the selected $\lambda_p$ is complex). In this way, based on the Popov-Belevitch-Hautus (PBH) test \cite{rugh1996linear}, the mode $\lambda_p$ of the system dynamics is made unobservable since $\mathbf{C}\hat{\mathbf{v}}_p=\mathbf{0}$. The design preserves most of the remaining eigenvectors and all the eigenvalues to maintain the open-loop eigenstructure as much as possible. 

In our previous study \cite{al2022observability}, we showed that we require two more actuation\footnote{If all the eigenvalues are real, then we would need one more actuation.} than the number of entries made zero in $\hat{\mathbf{v}}_p$. Thus we could modify $2m$ entries of $\mathbf{v}_p$ (i.e. $n-m+1, n-m+2, \cdots, n$ and $2n-m+1, 2n-m+2, \cdots, 2n$ entries) to become zero by using $q=2m+2$ actuation nodes\footnote{If all the eigenvalues are real, then we would need $q=2m+1$ actuation nodes.}. However, it is possible to achieve observability-blocking through $q=m+2$ actuation nodes, by exploiting the dynamics of DIN. To do so, we now present a result which establishes a relationship among entries of an eigenvector of the closed-loop DIN model.

\medskip

\noindent \textbf{Lemma 1:}  \textit{Consider the closed-loop dynamics of the DIN model given by (\ref{eq:main}). Suppose $\lambda_p$ and $\hat{\mathbf{v}}_p$ are the eigenvalue and eigenvector pair of $(\mathbf{A}+\mathbf{B}\mathbf{F})$. Then the $j$-th and $(j+n)$th entries of the eigenvector, denoted as $
\hat{{v}}_{p,j}$ and $\hat{{v}}_{p,j+n}$, are related by: $\hat{{v}}_{p,j+n}= \lambda_p~
\hat{v}_{p,j}$, where $j=1, 2, \cdots, n$.}

\medskip

\noindent \textbf{Proof:}

To establish the relationship among entries of $\hat{\mathbf{v}}_p$, we partition $\hat{\mathbf{v}}_p$ as $\hat{\mathbf{v}}_p= \begin{bmatrix}
    \hat{\mathbf{v}}_p^s \\
     \hat{\mathbf{v}}_p^{\dot{s}} \end{bmatrix}$ where both $\hat{\mathbf{v}}_p^s$ and  $\hat{\mathbf{v}}_p^{\dot{s}}$ are vectors in $\mathbb{C}^n$. Since $\mathbf{A} = \begin{bmatrix}
    \mathbf{0} ~~~~~~ \mathbf{I} \\ 
    -\mathbf{L}^s ~~ -\mathbf{L}^{\dot{s}} \end{bmatrix}$, $\mathbf{B} = \begin{bmatrix}
    \mathbf{0} \\
    \mathbf{\hat{B}}\end{bmatrix}$, we can write the equation $\mathbf{(A+BF)}\hat{\mathbf{v}}_p = \lambda_p \hat{\mathbf{v}}_p$ as

\begin{eqnarray} \label{eq:prf-le1}
    \left(\begin{bmatrix}
    \mathbf{0} & \mathbf{I} \\
    -\mathbf{L}^s & -\mathbf{L}^{\dot{s}}
\end{bmatrix}+
\begin{bmatrix}
    \mathbf{0} \\
    \mathbf{\hat{B}F}\end{bmatrix}\right)
\begin{bmatrix}
    \hat{\mathbf{v}}_p^s \\
     \hat{\mathbf{v}}_p^{\dot{s}}\end{bmatrix}= \lambda_p \begin{bmatrix}
    \hat{\mathbf{v}}_p^s \\
     \hat{\mathbf{v}}_p^{\dot{s}}\end{bmatrix}
\end{eqnarray}
From the first row of (\ref{eq:prf-le1}), we get the equation:
\begin{eqnarray} \label{eqn:pos-vel}
     \hat{\mathbf{v}}_p^{\dot{s}} = \lambda_p~\hat{\mathbf{v}}_p^s
\end{eqnarray}
From the partition of $\hat{\mathbf{v}}_p$ and eqn. (\ref{eqn:pos-vel}), it is immediate that $\hat{{v}}_{p,j+n}= \lambda_p~\hat{{v}}_{p,j}$ for $j=1, 2, \cdots, n$.

\hfill\(\Box\)

Note that Lemma 1 also holds for the open-loop dynamics of DIN, as seen by setting $\mathbf{F}=0$. Now, from Lemma 1 we see that if $\hat{{v}}_{p,j}=0$ then $\hat{{v}}_{p,j+n}=0$, where $j=1, 2, \cdots, n$. 
Thus, if we enforce zero to $n-m+1, n-m+2, \cdots, n$ entries of $\hat{\mathbf{v}}_p$, then the $2n-m+1, 2n-m+2, \cdots, 2n$ entries will also become zero. Therefore, we can adopt Algorithm 1 of \cite{al2022observability} and update it to enforce observability to our DIN model (\ref{eq:main}) by using only $q=m+2$ actuation nodes. The adapted algorithm for blocking observability in DIN based on \cite{al2022observability} is presented in the appendix. We formalize the result of the algorithm in the following theorem.

\medskip

\noindent \textbf{Theorem 1:}  \textit{Consider the DIN model \eqref{eq:main}. Assume that: (1) the eigenvalues of $\mathbf{A}$ are non-defective\footnote{Reference \cite{al2022observability} initially assumed distinct eigenvalues for development of Theorem 1, but later it was shown that Theorem 1 applies for non-defective eigenvalues also. We present this result for non-defective case as DIN is known to have repeated eigenvalues.}, (2) $q \geq m+2$, and (3) the pair $(\mathbf{A},\mathbf{B})$ is controllable. Then the gain matrix $\mathbf{F}$ of the state feedback controller can be designed to block observability of the model, i.e. to make the pair $(\mathbf{C},(\mathbf{A}+\mathbf{BF}))$ unobservable.  Specifically, the controller can be designed via Algorithm 1 so that any selected non-zero open-loop eigenvalue $\lambda_p$ becomes unobservable. Furthermore, all open-loop eigenvalues and the open-loop eigenvectors in the set $V_0 \cap V_1$, as defined in Algorithm 1, are maintained in the closed-loop model.}

\medskip

\noindent \textbf{Proof:}

Given that conditions (1)-(3) are satisfied, Theorem 1 of \cite{al2022observability} guarantees that the algorithm is always able to enforce $m$ entries of $\hat{\mathbf{v}}_p$ to become zero, while maintaining all open-loop eigenvalues and open-loop eigenvectors of the set $V_0 \cap V_1$. Now, since we take $\mathbf{N}_4(\lambda_p)~ \mathbf{h}_p = \mathbf{0}$ in Step 3, the algorithm enforces zero to the $n-m+1, n-m+2, \cdots, n$ entries of $\hat{\mathbf{v}}_p$. However, Lemma 1 implies that $2n-m+1, 2n-m+2, \cdots, 2n$ entries of $\hat{\mathbf{v}}_p$ will also be zero. Therefore, according to the (PBH) test, the eigenvalue $\lambda_p$ of the system dynamics is made unobservable since $\mathbf{C}\hat{\mathbf{v}}_p=\mathbf{0}$.

\hfill\(\Box\)

According to \cite{al2022observability}, we note that when all the eigenvalues are real, only $q=m+1$ actuation nodes are needed. We refer readers to \cite{al2022observability} for important remarks on Algorithm 1. We make the following additional remarks regarding Algorithm 1 presented here:

1) Algorithm 1 and Theorem 1 as presented here can be applied to the case when $\mathbf{A}$ has defective eigenvalues. Reference \cite{al2022observability} outlines the modifications needed for the case of defective eigenvalues, which involves the modification of generalized eigenvectors. The modified algorithm of \cite{al2022observability} can be adjusted following the same adjustment made in Algorithm 1 here to enforce observability in a DIN model. We omit further discussion on this for the sake of simplicity and to avoid repetition. 

2) In Step 5, we could choose $\mathbf{h}_p$ so that
$\mathbf{N}_6(\lambda_p)~ \mathbf{h}_p = \mathbf{0}$. In that case, the algorithm will directly enforce $2n-m+1, 2n-m+2, \cdots, 2n$ entries of $\hat{\mathbf{v}}_p$ to become zero. But according to Lemma 1, $n-m+1, n-m+2, \cdots, n$ entries of $\hat{\mathbf{v}}_p$ will also become zero given that $\lambda_p\neq 0$. Thus, unobservability will also be ensured with an additional condition that the selected eigenvalue is nonzero in Step 1 i.e. $\lambda_p\neq 0$. 

3) The result presented here does not depend on the Laplacian forms of $\mathbf{L}^s$ and $\mathbf{L}^{\dot{s}}$. Thus, the algorithm can be applied to any DIN which does not achieve synchronization or consensus for a zero input response.

\subsection{Sparser Observability-Blocking Using Network Cutsets}

In this section, we show that by utilizing the topological structure of DIN, we can induce unobservability with a smaller set of actuation nodes than the $q=m+2$ action nodes. This sparser design follows from the insight that blocking observability at the nodes associated with a cutset of the network graph using actuation in one partition can serve to block observability at all nodes associated with the other partition. We formalize this result first, and then provide our sparser observability-blocking design.

For the convenience of our development, as we did in \cite{al2019observability,al2022observability}, here we define two network models which have different measurement paradigms. We consider a {\em base network model} as defined in Section II, with defined state, input, and output matrices for the model as $\mathbf{A}$, $\mathbf{B}$, and $\mathbf{C}$.
A second model is then defined (with the same state dynamics and actuation nodes), but with a distinct measurement model. This measurement model considers a node-cutset of the graph $\mathcal{G}$ that partitions the actuation and measurement nodes in the base model. As shown in Fig. 1, the cutset $\mathcal{V}_{cut}$ separates the graph into two partitions $\mathcal{V}_1$ and $\mathcal{V}_2$  in such a way that $\mathcal{V}_1$ does not include any measurement nodes and $\mathcal{V}_2$ does not include any actuation nodes. We aptly name this second model the {\em cutset-measurement network model}. We stress that the cutset may include the actuation or measurement nodes of the base model as shown in Fig. 1, and hence a cutset always exists such that $|\mathcal{V}_{cut}| \leq m$ where $|\mathcal{V}_{cut}|$ represents the cardinality of $\mathcal{V}_{cut}$. In the cutset-measurement network model, the measurement nodes are defined as the nodes of this cutset. Therefore, the corresponding output matrix is given by $\tilde{\mathbf{C}}=\begin{bmatrix}
    \mathbf{\bar{C}} ~~ \mathbf{0}\\
    \mathbf{0} ~~ \mathbf{\bar{C}}\end{bmatrix}$ where $\mathbf{\bar{C}}\in \mathbb{R}^{|\mathcal{V}_{cut}| \times n}$.

Without loss of generality, the nodes are renumbered as follows: the nodes in $\mathcal{V}_1$ are ordered first, then the nodes in $\mathcal{V}_{cut}$, and lastly the nodes in $\mathcal{V}_2$. Therefore, the Laplacian $\mathbf{L}^s$ and $\mathbf{L}^{\dot{s}}$ can now be partitioned into blocks as 
\begin{eqnarray} \label{eq:par_L}
\mathbf{L}^s=
  \begin{bmatrix}
    \mathbf{L}^s_{11} & \mathbf{L}^s_{1c} & \mathbf{0} \\
    \mathbf{L}^s_{c1} & \mathbf{L}^s_{cc} & \mathbf{L}^s_{c2}\\
    \mathbf{0} & \mathbf{L}^s_{2c} & \mathbf{L}^s_{22} 
  \end{bmatrix},~~~~ \mathbf{L}^{\dot{s}}=\begin{bmatrix}
    \mathbf{L}^{\dot{s}}_{11} & \mathbf{L}^{\dot{s}}_{1c} & \mathbf{0} \\
    \mathbf{L}^{\dot{s}}_{c1} & \mathbf{L}^{\dot{s}}_{cc} & \mathbf{L}^{\dot{s}}_{c2}\\
    \mathbf{0} & \mathbf{L}^{\dot{s}}_{2c} & \mathbf{L}^{\dot{s}}_{22} 
  \end{bmatrix} 
\end{eqnarray}
where the block subscripts $1,  2$ and $c$ indicate the node sets $\mathcal{V}_1, \mathcal{V}_2$ and $\mathcal{V}_{cut}$. Thereby, $\mathbf{L}^s_{1c}$ refers to the block of $\mathbf{L}^s$ whose rows and columns correspond to the nodes of $\mathcal{V}_1$ and  $\mathcal{V}_{cut}$ respectively.

The following lemma gives a condition under which observability-blocking in the cutset-measurement network model implies observability-blocking in the base model. This lemma will then be used to obtain a sparser observability-blocking controller design.

\medskip

\noindent \textbf{Lemma 2:}  \textit{Consider the base network model \eqref{eq: open-loop} and its associated cutset-measurement network model. Suppose the gain matrix $\mathbf{F}$ of the state feedback controller is designed to block observability in the cutset-measurement network model, in such a way that the pair $(\hat{\mathbf{C}},(\mathbf{A}+\mathbf{BF}))$ has an unobservable eigenvalue at $\lambda_p$. Assume, $\lambda_p^2$ is not an eigenvalue of the matrix $\mathbf{L}_g$ where $\mathbf{L}_g=-(\mathbf{L}^s_{22} + \lambda_{p} \mathbf{L}^{\dot{s}}_{22})$. Then this feedback controller also serves to block observability in the base network model, specifically the pair $(\mathbf{C},(\mathbf{A}+\mathbf{BF}))$ also has an unobservable eigenvalue at $\lambda_p$.}

\medskip

\noindent \textbf{Proof:}

\begin{figure}[thpb] \label{fig2}
\centering
\includegraphics[width=8cm,height=4.5cm]{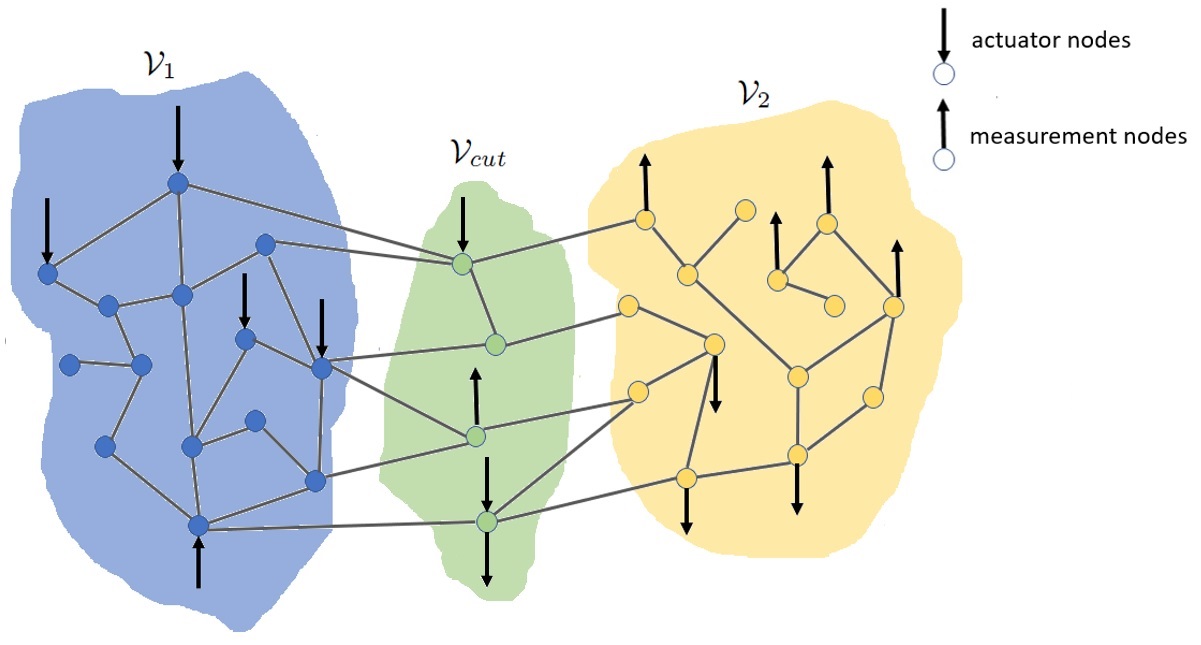}
\caption{Graph $\mathcal{G}$ of the network with node-cutset between actuation and measurement nodes.} \label{fig1}
\end{figure}

Consider the DIN model defined on graph $\mathcal{G}=(\mathcal{V},\mathcal{E})$, and the node-sets $\mathcal{V}_1, \mathcal{V}_{cut}$, and $\mathcal{V}_2$ as shown in Fig. 1. Note, the proof is immediate for the trivial case where $\mathcal{V}_2= \emptyset$, because in that case $\hat{\mathbf{C}}= \mathbf{C}$. Now consider that $\mathcal{V}_2 \neq \emptyset$. For this case, the $\mathbf{A}$ matrix can be partitioned as

\begin{eqnarray}
\resizebox{.9\hsize}{!}{
$\mathbf{A}=
  \begin{bmatrix}
    \mathbf{0} & \mathbf{0} & \mathbf{0}  & \mathbf{I} & \mathbf{0} & \mathbf{0}\\
    \mathbf{0} & \mathbf{0} & \mathbf{0}  & \mathbf{0} & \mathbf{I} & \mathbf{0}\\
    \mathbf{0} & \mathbf{0} & \mathbf{0}  & \mathbf{0} & \mathbf{0} & \mathbf{I}\\
    -\mathbf{L}^s_{11} & -\mathbf{L}^s_{1c} & \mathbf{0} & -\mathbf{L}^{\dot{s}}_{11} & -\mathbf{L}^{\dot{s}}_{1c} & \mathbf{0}  \\
    -\mathbf{L}^s_{c1} & -\mathbf{L}^s_{cc} & -\mathbf{L}^s_{c2} & -\mathbf{L}^{\dot{s}}_{c1} & -\mathbf{L}^{\dot{s}}_{cc} & -\mathbf{L}^{\dot{s}}_{c2} \\
    \mathbf{0} & -\mathbf{L}^s_{2c} & -\mathbf{L}^s_{22} & \mathbf{0} & -\mathbf{L}^{\dot{s}}_{2c} & -\mathbf{L}^{\dot{s}}_{22} 
  \end{bmatrix}$}
\end{eqnarray}

Assume $\lambda_p$ and $\mathbf{\hat{v}}_p$ are the unobservable eigenvalue and corresponding eigenvector of the cutset-measurement network model. We can then partition $\mathbf{\hat{v}}_p$ as $\mathbf{\hat{v}}_p=[(\mathbf{\hat{v}}^s_{p,1})^T ~ \mathbf{0} ^T~ (\mathbf{\hat{v}}^s_{p,2})^T  ~ (\mathbf{\hat{v}}^{\dot{s}}_{p,1})^T  ~ \mathbf{0}^T ~ (\mathbf{\hat{v}}^{\dot{s}}_{p,2})^T]^T$.
   
Now, we can re-write the equation $\mathbf{(A+BF)}\mathbf{\hat{v}}_p = \lambda_p \mathbf{\hat{v}}_p$ as
\begin{eqnarray} \label{eq6}
\resizebox{.95\hsize}{!}{
$  \begin{bmatrix}
    \mathbf{0} & \mathbf{0} & \mathbf{0}  & \mathbf{I} & \mathbf{0} & \mathbf{0}\\
    \mathbf{0} & \mathbf{0} & \mathbf{0}  & \mathbf{0} & \mathbf{I} & \mathbf{0}\\
    \mathbf{0} & \mathbf{0} & \mathbf{0}  & \mathbf{0} & \mathbf{0} & \mathbf{I}\\
    -\mathbf{L}^s_{11}+\mathbf{u}^s_{11}  & -\mathbf{L}^s_{1c}+\mathbf{u}^s_{1c} & \mathbf{u}^s_{12} & -\mathbf{L}^{\dot{s}}_{11}+\mathbf{u}^{\dot{s}}_{11} & -\mathbf{L}^{\dot{s}}_{1c}+\mathbf{u}^{\dot{s}}_{1c}& \mathbf{u}^{\dot{s}}_{12}  \\
    -\mathbf{L}^s_{c1}+\mathbf{u}^s_{c1}& -\mathbf{L}^s_{cc}+\mathbf{u}^s_{cc} & -\mathbf{L}^s_{c2}+\mathbf{u}^s_{c2} & -\mathbf{L}^{\dot{s}}_{c1}+\mathbf{u}^{\dot{s}}_{c1} & -\mathbf{L}^{\dot{s}}_{cc}+\mathbf{u}^{\dot{s}}_{cc} & -\mathbf{L}^{\dot{s}}_{c2}+\mathbf{u}^{\dot{s}}_{c2} \\
    \mathbf{0} & -\mathbf{L}^s_{2c} & -\mathbf{L}^s_{22} & \mathbf{0} & -\mathbf{L}^{\dot{s}}_{2c} & -\mathbf{L}^{\dot{s}}_{22} 
  \end{bmatrix}
 \begin{bmatrix}
    \mathbf{\hat{v}}^s_{p,1}  \\
   \mathbf{0} \\
   \mathbf{\hat{v}}^s_{p,2}  \\
   \mathbf{\hat{v}}^{\dot{s}}_{p,1}  \\
   \mathbf{0} \\
   \mathbf{\hat{v}}^{\dot{s}}_{p,2}  
\end{bmatrix}$}\nonumber
  \end{eqnarray} 

\begin{eqnarray}
= \lambda_p
\begin{bmatrix}
    \mathbf{\hat{v}}^s_{p,1}  \\
   \mathbf{0} \\
   \mathbf{\hat{v}}^s_{p,2}  \\
   \mathbf{\hat{v}}^{\dot{s}}_{p,1}  \\
   \mathbf{0} \\
   \mathbf{\hat{v}}^{\dot{s}}_{p,2}  
\end{bmatrix}
  \end{eqnarray}
  
Here, $\mathbf{u}^s_{12}$ refers to the actuation on the nodes in $\mathcal{V}_1$ exerted by the nodes in $\mathcal{V}_2$ through feedback and edge-weight set $\mathcal{W}^s$. Now, from the last row of (\ref{eq6}) we obtain the following equation: 
\begin{eqnarray} \label{eq7}
    -\mathbf{L}^s_{22} \mathbf{\hat{v}}^s_{p,2} - \mathbf{L}^{\dot{s}}_{22} \mathbf{\hat{v}}^{\dot{s}}_{p,2} &=& \lambda_p \mathbf{\hat{v}}^{\dot{s}}_{p,2}
\end{eqnarray}
From Lemma 1, we know
\begin{eqnarray} \label{eq8}
\mathbf{\hat{v}}^{\dot{s}}_{p,2} &=& \lambda_p \mathbf{\hat{v}}^s_{p,2}
\end{eqnarray}
Using (\ref{eq8}), we can write (\ref{eq7}) as:
\begin{eqnarray} \label{eq9}
    \lambda_{p}^2 \mathbf{\hat{v}}^s_{p,2} = -(\mathbf{L}^s_{22} + \lambda_{p} \mathbf{L}^{\dot{s}}_{22}) \mathbf{\hat{v}}^s_{p,2}= \mathbf{L}_g \mathbf{\hat{v}}^s_{p,2}
\end{eqnarray}
where $\mathbf{L}_g=-(\mathbf{L}^s_{22} + \lambda_{p} \mathbf{L}^{\dot{s}}_{22})$. If $\lambda_p^2$ is not an eigenvalue of $\mathbf{L}_g$, then (\ref{eq9}) implies that $\mathbf{\hat{v}}^s_{p,2}=\mathbf{0}$. Then according to (\ref{eq8}), we get $\mathbf{\hat{v}}^{\dot{s}}_{p,2} =\mathbf{0}$. Therefore, $\mathbf{\hat{v}}_p$ can be written as $\mathbf{\hat{v}}_p=[(\mathbf{\hat{v}}^s_{p,1})^T ~ \mathbf{0} ^T~ \mathbf{0} ^T  ~ (\mathbf{\hat{v}}^{\dot{s}}_{p,1})^T  ~ \mathbf{0}^T ~ \mathbf{0} ^T]^T$. Note, $\mathcal{V}_2 \cup \mathcal{V}_{cut}$ contains all the measurement nodes and the entries of $\hat{\mathbf{v}}_p$ corresponding to the nodes in $\mathcal{V}_2 \cup \mathcal{V}_{cut}$ are zero. Hence, the proof is complete.

\hfill\(\Box\)

\medskip

Note, the condition given in Lemma 2 is completely different than what we had in our earlier work \cite{al2022observability}. Now, Lemma 2 serves as a basis for our algorithm for sparser observability-blocking control. The idea is to first use the algorithm from Section III-A to design an observability-blocking controller for the cutset-measurement network model, whereupon Lemma 2 can be leveraged to guarantee that observability is blocked in the base model also. The overall design procedure can be described as follows: First, a small-cardinality vertex-cutset $\mathcal{V}_{cut}$ separating the measurement nodes from the actuation nodes in the network graph is chosen, and the associated cutset-measurement network model is formed. Then, in accordance with the algorithm in Section III-A, an eigenvalue $\lambda_p$ of $\mathbf{A}$ is chosen, whose eigenvector will be modified to block observability. However, an additional condition is imposed on the selection, that $\lambda_p^2$ is not an eigenvalue of $\mathbf{L}_g=-(\mathbf{L}^s_{\mathcal{V}_2\mathcal{V}_2} + \lambda_{p} \mathbf{L}^{\dot{s}}_{\mathcal{V}_2\mathcal{V}_2})$. Thereupon, the algorithm presented in Section III-A can be applied to find the controller that makes the eigenvalue at $\lambda_p$ unobservable in the cutset-measurement network model using $|\mathcal{V}_{cut}|+2$ actuation nodes. From Lemma 2, it is immediate that the same controller blocks observability in the base network model. In this way an observability-blocking controller is found that requires only $|\mathcal{V}_{cut}|+2$ actuation nodes. Further, this controller maintains all of the open-loop eigenvalues, according to Theorem 1. We formalize this design in the following theorem. 

\medskip
 
\noindent \textbf{Theorem 2:} \textit{Consider the network model \eqref{eq: open-loop}, and say that a separating vertex-cutset $V_{cut}$ has been found. Assume that: 1) $q \geq |\mathcal{V}_{cut}|+2$, 2) the graph $\mathcal{G}$ of the network model is strongly connected and has positive edge-weights, and 3) the pair ($\mathbf{A},\mathbf{B})$ is controllable. Then the gain matrix $\mathbf{F}$ of state feedback controller can be designed to block observability, i.e. to make the pair $(\mathbf{C},(\mathbf{A}+\mathbf{BF}))$ unobservable. Furthermore, all the open-loop eigenvalues are maintained in the closed-loop system for the design.}

\medskip

\noindent \textbf{Proof:}

For the proof, all we need to show is that there exists an eigenvalue $\lambda_p$ of $\mathbf{A} \in \mathbb{R}^{2n\times 2n}$ such that $\lambda_p^2$ is not an eigenvalue of $\mathbf{L}_g$. To show that, we choose a fixed eigenvalue of $\mathbf{A}$, which is $0$. Now, choosing $\lambda_p=0$, we get $\mathbf{L}_g=-\mathbf{L}^s_{22}$. Since the network graph $\mathcal{G}$ is strongly connected and the edge-weights of the graph are assumed to be positive, then the grounded Laplacian $\mathbf{L}^s_{22}$ has all of its eigenvalues in the open right half plane (see Theorem 1 of \cite{xia2017analysis}). Hence, $\mathbf{L}_g$ has all of its eigenvalues in the open left half plane and cannot have a zero eigenvalue. Therefore, we can conclude that there always exists an eigenvalue $\lambda_p$ to choose so that the condition mentioned in Lemma 2 is satisfied. 

\hfill\(\Box\)

We can see that Theorem 2 provides an appealing approach for enforcing unobservability in a sparse DIN network with a small cutset. Many real-world networks, such as power-grids, transportation, and scale-free networks, are sparse and have small-cardinality cutsets \cite{del2011all}. In such networks, sparser design is useful to block observability  using only a few actuator nodes. Note, when all the eigenvalues of $\mathbf{A}$ are real, we can use $q = |\mathcal{V}_{cut}|+1$ actuation nodes in our design. We further remark that the results presented here do not rely on the Laplacian forms of $\mathbf{L}^s$ and $\mathbf{L}^{\dot{s}}$ matrices, except when there is need for the existence of a zero eigenvalue in the proof of Theorem 2. Given that the selected eigenvalue satisfies the technical criterion: $\lambda_p^2$ is not an eigenvalue of matrix $\mathbf{L}_g$, the sparser design holds for any form of $\mathbf{L}^s$ and $\mathbf{L}^{\dot{s}}$ matrices.

\subsection{Extension to $N$-th Order Integrator Network}

Following the approach presented in this paper, we can easily extend our results for an $N$-th order integrator network where $N \in \mathbb{N}$. In this section we present the results and briefly discuss their development. We omit formal proofs as these can be easily derived using the same arguments presented in previous two sections of this paper.

In an $N$-th order integrator network, each node $i$ is associated with $N$ states, which can be represented as $s_i(t), \dot{s}_i(t), \ddot{s}_i(t), \cdots, s^{N-1}_i(t)$. Here, $s^{N-1}_i(t)$ denotes the $(N-1)$-th  time-derivative of $s_i(t)$. The network states are defined as $\mathbf{x}(t)= [s_1(t) ~ \cdots ~ s_n(t) ~\cdots ~ s_1^{N-1}(t) ~ \cdots ~ s_n^{N-1}(t)]^T$. We again enhance this network model by capturing a set of actuation nodes and a set of measurement nodes, similar to Section II. The $N$-th order integrator network dynamics with actuation and measurement included, is then given by the following state space model:

\begin{subequations} \label{eq:dynamics_NIN}
\begin{align}
\mathbf{\dot{x}} =&  \begin{bmatrix}
    \mathbf{0} ~~~~\mathbf{I}~~~~\mathbf{0}~~~~ \cdots~~~~ \mathbf{0} \\ 
    \mathbf{0} ~~~~\mathbf{0}~~~~ \mathbf{I}~~~~ \cdots~~~~ \mathbf{0} \\
    \vdots ~~~~ \vdots ~~~~~ \vdots ~~~~~ \vdots ~~~~~ \vdots \\
    \mathbf{0} ~~~~\mathbf{0}~~~~ \mathbf{0}~~~~\cdots~~~~ \mathbf{I} \\
    -\mathbf{L}^s ~ -\mathbf{L}^{\dot{s}}~ -\mathbf{L}^{\ddot{s}} \cdots~ -\mathbf{L}^{s^{N-1}} \end{bmatrix} \mathbf{x} + \begin{bmatrix}
    \mathbf{0} \\
    \mathbf{0} \\
    \vdots \\
    \mathbf{0}\\
    \mathbf{\hat{B}}\end{bmatrix}  \mathbf{u}, \\
\mathbf{y} =& \begin{bmatrix}
    \mathbf{\hat{C}} ~~ \mathbf{0}  \cdots \mathbf{0} \\
    \mathbf{0} ~~ \mathbf{\hat{C}} \cdots \mathbf{0}\\
    \vdots ~~~\vdots ~~~\vdots ~~~ \vdots \\
    \mathbf{0} ~~ \mathbf{0} \cdots \mathbf{\hat{C}}\\ 
\end{bmatrix} \mathbf{x}
\end{align}
\end{subequations}
where $\mathbf{\hat{B}}$ and $\mathbf{\hat{C}}$ are defined as before. Therefore, the open-loop dynamics and closed-loop dynamics of the model can be written as (\ref{eq: open-loop}) and (\ref{eq:main}) respectively. Again, our objective is to design a state-feedback controller, defined by the gain matrix $\mathbf{F}$, to enforce that the pair $(\mathbf{C},(\mathbf{A}+\mathbf{BF}))$ is unobservable, while maintaining as much of the open-loop eigenstructure as possible.  

The block structure of the state matrix $\mathbf{A}$ in (\ref{eq:dynamics_NIN}) allows us to generalize Lemma 1 for the $N$-th order integrator network as below.

\medskip

\noindent \textbf{Lemma 3:}  \textit{Consider the closed-loop dynamics (\ref{eq:main}) of an $N$-th order integrator network model. Suppose $\lambda_p$ and $\hat{\mathbf{v}}_p$ are the eigenvalue and eigenvector pair of $(\mathbf{A}+\mathbf{B}\mathbf{F})$. Then the entries of the eigenvector are related as: $\hat{{v}}_{p,j+(N-1)n}= \lambda_p~
\hat{v}_{p,j+(N-1)n}= \cdots =\lambda_p^{N-1}~
\hat{v}_{p,j}$, where $j=1, 2, \cdots, n$.}

\medskip

Therefore, Lemma 3 implies that if $\hat{{v}}_{p,j}=0$, where $j=1, 2, \cdots, n$, then $\hat{{v}}_{p,j+n}=\hat{{v}}_{p,j+2n}= \cdots =\hat{{v}}_{p,j+(N-1)n}=0$.  Thus, if we enforce zero to $n-m+1, n-m+2, \cdots, n$ entries of $\hat{\mathbf{v}}_p$, other $m(N-1)$ entries of $\hat{\mathbf{v}}_p$ will become zero. Therefore we can use Algorithm 1 and update it according to the $N$-th order integrator network model to enforce unobservability. We skip providing the algorithm for this case to avoid repetition. Then Theorem 1 can also be generalized as below, based on Lemma 3.

\medskip

\noindent \textbf{Theorem 3:}  \textit{Consider an $N$-th order integrator network model. Assume that: 1) the eigenvalues of $\mathbf{A}$ are non-defective, 2) $q \geq m+2$, and 3) the pair $(\mathbf{A},\mathbf{B})$ is controllable. Then the gain matrix $\mathbf{F}$ of the state feedback controller can be designed to block observability of the model, i.e. to make the pair $(\mathbf{C},(\mathbf{A}+\mathbf{BF}))$ unobservable.  Specifically, the controller can be designed using the update of Algorithm 1 so that any selected non-zero open-loop eigenvalue $\lambda_p$ becomes unobservable. Furthermore, all open-loop eigenvalues and the open-loop eigenvectors in the set $V_0 \cap V_1$, as defined in Algorithm 1, are maintained in the closed-loop model.}

\medskip

Now we provide the result for sparser design. For the sparser design, we again define a cutset-measurement network model similar to Section III-B. Without loss of generality, the nodes can again be renumbered as follows: the nodes in $\mathcal{V}_1$ are ordered first, then nodes in $\mathcal{V}_{cut}$, and lastly nodes in $\mathcal{V}_2$. This would allow us to partition the Laplacian matrices $\mathbf{L}^s, \mathbf{L}^{\dot{s}}, \cdots~ \mathbf{L}^{s^{N-1}}$ in the same way as (\ref{eq:par_L}). Then using the approach presented in the proof of Lemma 2, we can establish the following results for the $N$-th order integrator network.

\noindent \textbf{Lemma 4:}  \textit{Consider an $N$-th order integrator network model and it's associated cutset-measurement network model. Suppose the gain matrix $\mathbf{F}$ of the state feedback controller is designed to block observability in the cutset-measurement network model, in such a way that the pair $(\hat{\mathbf{C}},(\mathbf{A}+\mathbf{BF}))$ has an unobservable eigenvalue at $\lambda_p$. Assume, $\lambda_p^N$ is not an eigenvalue of the matrix $\mathbf{L}_g$, where $\mathbf{L}_g=-(\mathbf{L}^s_{22} + \lambda_{p} \mathbf{L}^{\dot{s}}_{22}+ \cdots + \lambda_{p}^{N-1} \mathbf{L}^{s^{N-1}}_{22})$. Then this feedback controller also serves to block observability in the base network model, specifically the pair $(\mathbf{C},(\mathbf{A}+\mathbf{BF}))$ has an unobservable eigenvalue at $\lambda_p$.}

\medskip

Note that the above $\mathbf{L}^s_{22}, \mathbf{L}^{\dot{s}}_{22}, \cdots, \mathbf{L}^{s^{N-1}}_{22}$ denotes the block matrices associated with partition $\mathcal{V}_2$  for the Laplacian matrices $\mathbf{L}^s, \mathbf{L}^{\dot{s}}, \cdots,\mathbf{L}^{s^{N-1}}$. Again, we can argue that $\mathbf{A}$ always has such an eigenvalue that satisfies the condition in Lemma 4. For example, $\lambda_p=0$ is an eigenvalue of $\mathbf{A}$, but is not an eigenvalue of $\mathbf{L}_g$ (based on the logic presented in the proof of Theorem 2). Therefore, we can generalize Theorem 2 for $N$-th order integrator network as below.

\medskip
 
\noindent \textbf{Theorem 4:} \textit{Consider an $N$-th order integrator network model, and suppose that a separating vertex-cutset $V_{cut}$ has been found. Assume that: 1) $q \geq |\mathcal{V}_{cut}|+2$, 2) the graph $\mathcal{G}$ of the network model is strongly connected and has positive edge-weights, and 3) the pair ($\mathbf{A},\mathbf{B})$ is controllable. Then the gain matrix $\mathbf{F}$ of the state feedback controller can be designed to block observability, i.e. to make the pair $(\mathbf{C},(\mathbf{A}+\mathbf{BF}))$ unobservable. Furthermore, all the open-loop eigenvalues are maintained in the closed-loop system for the design.}

\medskip

Note, when all the eigenvalues of $\mathbf{A}$ are real, we can use $q = m+1$ or $q = |\mathcal{V}_{cut}|+1$ actuation nodes in Theorem 3 and Theorem 4 respectively. Also, the results does not rely on the Laplacian forms of $\mathbf{L}^s, \mathbf{L}^{\dot{s}}, \cdots,\mathbf{L}^{s^{N-1}}$ matrices, given that the condition on $\lambda_p$ mentioned in Lemma 4 is satisfied.

\section{Numerical Example}:

\begin{figure}[thpb] 
\centering
\includegraphics[width=7cm,height=3.5cm]{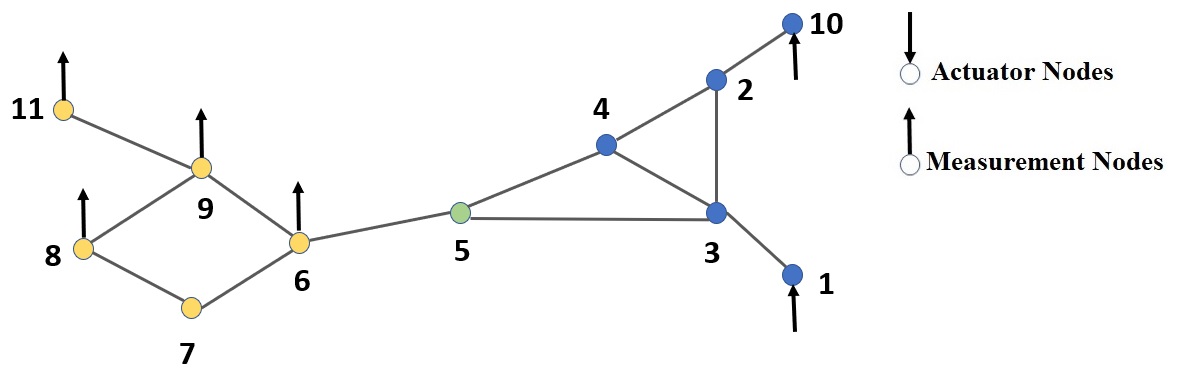}
\caption{Network graph with double-integrator dynamics used in the numerical example} 
\end{figure} \label{fig3}

In this section, we present a numerical example that verifies our theoretical findings. We adopt the same example graph used in \cite{al2022observability}, but now considering DIN dynamics for the nodes. Specifically, an $11$ node network with an undirected graph $\mathcal{G}$ is used as an example, as shown in Fig. 2. The edge weights for  $\mathcal{G}$ in the Laplacians $\mathbf{L}^s$ and $\mathbf{L}^{\dot{s}}$ are assigned random positive numbers.
The measurement nodes are assumed to be $\{ 6, 8, 9, 11 \}$. We identify from Fig. 2 that node 5 is a potential single-node cutset that results in the graph's measurement nodes coalesced into one partition. Thus, from Theorems 1 and 2, we should be able to build state feedback controllers at any three nodes from $\{1, 2, 3, 4, 10\}$ to block observability at node 5, and hence at the measurement nodes. However, since all the eigenvalues of $\mathbf{A}$ are real, $q=2$ is sufficient to block observability by using our algorithm. We choose nodes 1 and 10 for actuation.

To build the controller, we select one eigenvalue and its associated eigenvector of the matrix $\mathbf{A}$. Specifically we select $\lambda_p=-1.0099$ whose corresponding eigenvector is $\mathbf{v}_p = [0.0305, 0.1342, -0.1163, -0.0948, 0.0118, 0.4642, -0.4413,$ $ -0.0857, 0.1786, -0.0594, -0.0218, -0.0308, -0.1355,$ $ 0.1174, 0.0958, -0.0119, -0.4688, 0.4457, 0.0865, -0.1803,$ $ 0.0600, 0.0220]^T$. Then we build state feedback controllers at nodes 1 and 10 according to our proposed algorithm to block observability at node 5. The control gains for controllers at nodes 1 and 10 are obtained as  $[-0.2550,-1.1694,0.9500,0.5459,0.3783,-2.8561,2.1917,$ $-0.1588,-0.3110,0.3363,0.3482,-0.0535,-0.2354,$ $0.2040,0.1664,-0.0207,-0.8147,0.7745,0.1504,-0.3134,$ $0.1042,0.0382]^T$ and  $[-0.1309,-0.6005,0.4878,0.2803,$ $0.1943,-1.4666,1.1254,-0.0816,-0.1597,0.1727,0.1788,$ $-0.0275,-0.1209,0.1048,0.0854,-0.0106,-0.4183,$ $0.3977,0.0772,-0.1609,0.0535,0.0196,]^T$respectively. 
After applying these controllers, the modified eigenvector $\mathbf{\hat{v}}_p$ is obtained as $[-0.0046, 0.0033, -0.0021, 0.0017, 0, 0, 0, 0, 0, -0.0045, 0, $ $0.0046, -0.0033, 0.0022, -0.0017, 0, 0, 0, 0, 0, 0.0045, 0]^T$. We see that the entries corresponding to node $5$ as well as all the nodes in the partition $\mathcal{V}_2$, i.e. the node set $\{ 6,7,8,9, 11\}$, are zero. Thus, the DIN model has become unobservable at the measurement nodes, verifying our theorems. 

\setlength{\parskip}{3pt}
\section{Conclusions}

A design algorithm has been developed for observability-blocking controls in a double integrator network (DIN). The topological structure of DIN has then been exploited to reduce the number of actuation nodes needed to block observability. We have generalized the observability-blocking design for $N$-th order integrator networks. A numerical example has been presented to verify our findings. One interesting future direction for this study includes developing design of observability-blocking controls for more complex networks, such as nonlinear dynamical networks.

\section*{Acknowledgment}
We thank Prof. Sandip Roy at Texas A\&M University for his valuable suggestions and insights regarding this work.  

\section{Appendix}

\noindent \textbf{Algorithm 1: Observability blocking controls for DIN} 

\begin{enumerate}
\item [1)]  Select one eigenvalue $\lambda_p$ of $\mathbf{A}$ and its associated eigenvector $\mathbf{v}_p$, where $p \in \{1, 2, \hdots, 2n\}$. If $\lambda_p$ is real, follow the steps under Sub-Algorithm 1 to obtain the observability-blocking controller. Otherwise follow the steps under Sub-Algorithm 2.
\end{enumerate}

\textbf{Sub-Algorithm 1: }

\begin{enumerate}
\item [2)] Compute a matrix $\mathbf{N}(\lambda_p) \in \mathbb{C}^{(2n+q) \times q}$, whose columns are linearly independent and span the null space of $\mathbf{S}(\lambda_p)=[(\mathbf{A}-\lambda_p\mathbf{I})~~ \mathbf{B}]$.  Then partition $\mathbf{N}(\lambda_p)$ as $\mathbf{N}(\lambda_p)=[\mathbf{N}_1(\lambda_p)^T ~ \mathbf{N}_2(\lambda_p)^T]^T$, where  $\mathbf{N}_1(\lambda_p) \in \mathbb{C}^{2n \times q}$ and  $\mathbf{N}_2(\lambda_p) \in \mathbb{C}^{q \times q}$. Therefore $\mathbf{N}_1(\lambda_p)$ and $\mathbf{N}_2(\lambda_p)$ satisfy:
\begin{eqnarray}
[(\mathbf{A}-\lambda_p~\mathbf{I})~~ \mathbf{B}] \left[\begin{array}{c}
                                         \mathbf{N}_1(\lambda_p)   \\
                                         \mathbf{N}_2(\lambda_p)
                                          \end{array}\right]  = \mathbf{0}.  \label{eq4}
\end{eqnarray}

\item [3)] Partition $\mathbf{N}_1(\lambda_p)$ as
$\mathbf{N}_1(\lambda_p)=[\mathbf{N}_3(\lambda_p)^T ~ \mathbf{N}_4(\lambda_p)^T ~ \mathbf{N}_5(\lambda_p)^T ~ \mathbf{N}_6(\lambda_p)^T]^T$, where  $\mathbf{N}_3(\lambda_p) \in \mathbb{C}^{(n-m) \times q}$, $\mathbf{N}_4(\lambda_p) \in \mathbb{C}^{m \times q}$, $\mathbf{N}_5(\lambda_p) \in \mathbb{C}^{(n-m) \times q}$ and  $\mathbf{N}_6(\lambda_p) \in \mathbb{C}^{m \times q}$.  Then find a vector $\mathbf{h}_p \neq \mathbf{0}$ which lies in the null space of $\mathbf{N}_4(\lambda_p)$, i.e. which satisfies:  
\begin{eqnarray}
\mathbf{N}_4(\lambda_p)~ \mathbf{h}_p &=& \mathbf{0}. \label{eq2}
\end{eqnarray}

\item [4)] Compute the vectors $\hat{\mathbf{v}_p}$ and $\mathbf{z}_p$ as:
\begin{eqnarray}
\hat{\mathbf{v}}_p &=& \mathbf{N}_1(\lambda_p)~ \mathbf{h}_p \label{eq1} \\
\mathbf{z}_p &=& \mathbf{N}_2(\lambda_p)~ \mathbf{h}_p. \label{eq5}
\end{eqnarray}

\item [5)] Form the open-loop modal matrix $\mathbf{V}_0 \in \mathbb{C}^{2n \times 2n}$ (i.e. $\mathbf{V}_0=[ \mathbf{v}_1 ~ \mathbf{v}_2 ~ \cdots ~ \mathbf{v}_{2n}]$), and $q \times {2n}$ zero matrix $\mathbf{Z}_0$. Then check whether the $2n$ vectors in $\hat{V}_0 = V_0 \cup \{\hat{\mathbf{v}}_p \} \backslash \{ \mathbf{v}_p \}$ are linearly independent, where $V_0$ is a vector set containing all the open-loop eigenvectors. If yes, then follow Step 6 below to construct the closed-loop modal matrix $\mathbf{V}$ and associated $\mathbf{Z}$ matrix. Otherwise, skip Step 6, and follow Steps 7 through 9 to construct $\mathbf{V}$ and $\mathbf{Z}$ matrices. 

\item [6)] Construct the matrix $\mathbf{V} \in \mathbb{C}^{2n \times 2n}$ from $\mathbf{V}_0$ by replacing the column containing $\mathbf{v}_p$ with $\hat{\mathbf{v}}_p$. Similarly, construct the matrix $\mathbf{Z} \in \mathbb{C}^{q \times 2n}$ from $\mathbf{Z}_0$ by replacing the corresponding column with $\mathbf{z}_p$. Therefore, $\mathbf{V}=[ \mathbf{v}_1 \cdots \mathbf{v}_{p-1}~ \hat{\mathbf{v}}_p~\mathbf{v}_{p+1}\cdots \mathbf{v}_{2n}]$ and $\mathbf{Z}= [\mathbf{0} \cdots \mathbf{0} ~ \mathbf{z}_p~ \mathbf{0} \cdots \mathbf{0}]$.  Then jump to Step 10.

\item [7)] Find the largest-cardinality subset $V_1$ of $\hat{V}_0$ such that $\hat{\mathbf{v}}_p \in V_1$ and $V_1$ is a self-conjugate set of linearly independent vectors. 

\item [8)] Find $\hat{\mathbf{v}}_k$ in the column space of $\mathbf{N}_1(\lambda_k)$ for all $\mathbf{v}_k \in \hat{V}_0 \backslash V_1$ such that the set $\hat{V}= V_1 \cup \{\hat{\mathbf{v}}_k | \mathbf{v}_k \in \hat{V}_0 \backslash V_1 \}$ is a self-conjugate set of $2n$ linearly independent vectors. While doing so, maintain $\hat{\mathbf{v}}_{k_2}= \bar{\hat{\mathbf{v}}}_{k_1}$ whenever $\mathbf{v}_{k_2}= \bar{\mathbf{v}}_{k_1}$ and $\mathbf{v}_{k_1}, \mathbf{v}_{k_2} \in \hat{V}_0 \backslash V_1$. Next, find the corresponding $\mathbf{z}_k$ for all $\mathbf{v}_k \in \hat{V}_0 \backslash V_1$ such that $\mathbf{z}_k = \mathbf{N}_2(\lambda_k)~ \mathbf{h}_k$, where $\mathbf{h}_k$ solves $\hat{\mathbf{v}}_k = \mathbf{N}_1(\lambda_k)~ \mathbf{h}_k$. Note that $\mathbf{z}_{k_2}= \bar{\mathbf{z}}_{k_1}$ whenever $\mathbf{v}_{k_2}= \bar{\mathbf{v}}_{k_1}$. 

\item [9)] Construct $\mathbf{V}$ from $\mathbf{V}_0$ by replacing the columns containing $\mathbf{v}_p$ and all $\mathbf{v}_k\in \hat{V}_0 \backslash V_1$ with $\hat{\mathbf{v}}_p$ and corresponding $\hat{\mathbf{v}}_k$ respectively. In the same manner construct $\mathbf{Z}$ from $\mathbf{Z}_0$ by replacing the corresponding columns of $\mathbf{Z}_0$ with $\mathbf{z}_p$ and all $\mathbf{z}_k$ obtained in Step 8 respectively. 

\item [10)] Finally the gain matrix $\mathbf{F}$ for the observability-blocking controller is obtained by (\ref{eq3a}):
\begin{eqnarray}
 \mathbf{F} = \mathbf{Z}~\mathbf{V}^{-1}.  \label{eq3a}
\end{eqnarray}
\end{enumerate}

\textbf{Sub-Algorithm 2:} 

\begin{enumerate} 
\item [2-4)] Steps 2 through 4 remain exactly the same as for Sub-Algorithm 1. Additionally in Step 4, it is necessary to obtain $\bar{\hat{\mathbf{v}}}_p$ and the associated $\bar{\mathbf{z}}_{p}$ by taking complex conjugates of $\hat{\mathbf{v}}_{p}$ and $\mathbf{z}_{p}$, respectively. 

\item [5)] Form the open-loop modal matrix $\mathbf{V}_0 \in \mathbb{C}^{2n \times 2n}$ and $q \times 2n$ zero matrix $\mathbf{Z}_0$. Then check whether the $2n$ vectors in $\hat{V}_0 = V_0 \cup \{\hat{\mathbf{v}}_p, \bar{\hat{\mathbf{v}}}_p\} \backslash \{ \mathbf{v}_p, \bar{\mathbf{v}}_p  \}$ are linearly independent. If yes, follow Step 6; otherwise skip Step 6 and follow Steps 7 through 9 to find $\mathbf{V}$ and $\mathbf{Z}$. 

\item [6)] Construct the matrix $\mathbf{V}$ from $\mathbf{V}_0$ by replacing the columns having $\mathbf{v}_p$ and $\bar{\mathbf{v}}_p$ with $\hat{\mathbf{v}}_p$ and $\bar{\hat{\mathbf{v}}}_p$ respectively. Similarly, construct the matrix $\mathbf{Z}$ from $\mathbf{Z}_0$ by replacing the corresponding columns of $\mathbf{Z}_0$ with $\mathbf{z}_p$ and $\bar{\mathbf{z}}_p$ respectively. Then jump to Step 10.

\item [7)] Find the largest-cardinality subset $V_1$ of $\hat{V}_0$ such that $\hat{\mathbf{v}}_p$, $\bar{\hat{\mathbf{v}}}_p \in V_1$ and $V_1$ is a self-conjugate set of linearly independent vectors.

\item [8)] This step is the same as the Step 8 of Sub-Algorithm 1. Thus, find $\hat{\mathbf{v}}_k$ in the column space of $\mathbf{N}_1(\lambda_k)$ for all $\mathbf{v}_k \in \hat{V}_0 \backslash V_1$ such that the set $\hat{V}= V_1 \cup \{\hat{\mathbf{v}}_k | \mathbf{v}_k \in \hat{V}_0 \backslash V_1 \}$ is a self-conjugate set of $2n$ linearly independent vectors, and find all the corresponding $\mathbf{z}_k$. 

\item [9)] Construct $\mathbf{V}$ from $\mathbf{V}_0$ by replacing the columns containing $\mathbf{v}_p$, $\bar{\mathbf{v}}_p$  and all $\mathbf{v}_k\in \hat{V}_0 \backslash V_1$ with $\hat{\mathbf{v}}_p$, $\bar{\hat{\mathbf{v}}}_p$ and corresponding $\hat{\mathbf{v}}_k$ respectively. In the same manner, construct $\mathbf{Z}$ from $\mathbf{Z}_0$ by replacing the corresponding columns of $\mathbf{Z}_0$ with  $\mathbf{z}_p$, $\bar{\mathbf{z}}_p$ and all $\mathbf{z}_k$ obtained in Step 8 respectively.  

\item [10)] Like Sub-Algorithm 1, compute the gain matrix $\mathbf{F}$ using (\ref{eq3a}).
\end{enumerate}

	\bibliographystyle{ieeetr}
	\bibliography{references}

\end{document}